\documentclass[]{pasj00}%{emulateapj}%[onecolumn]{aastex}
\usepackage{natbib}
\usepackage{times}

\usepackage{url}

\begin{document}
\title{Metal Enrichment in the Fermi Bubbles as a Probe of Their Origin} 

\author{Yoshiyuki Inoue\altaffilmark{1}, Shinya Nakashima\altaffilmark{1}, Masaya Tahara\altaffilmark{2}, Jun Kataoka\altaffilmark{2}, Tomonori Totani\altaffilmark{3}, Yutaka Fujita\altaffilmark{4}, and Yoshiaki Sofue\altaffilmark{5}} 
\affil{$^1$Institute of Space and Astronautical Science JAXA, 3-1-1 Yoshinodai, Chuo-ku, Sagamihara, Kanagawa 252-5210, Japan}
\affil{$^2$Research Institute for Science and Engineering, Waseda University, 3-4-1, Okubo, Shinjuku, Tokyo 169-8555, Japan}
\affil{$^3$Department of Astronomy, The University of Tokyo, Bunkyo-ku, Tokyo 113-0033, Japan}
\affil{$^4$Department of Earth and Space Science, Graduate School of Science, Osaka University, Toyonaka, Osaka 560-0043, Japan}
\affil{$^5$Institute of Astronomy, The University of Tokyo, Mitaka, Tokyo, 181-0015}
\email{E-mail: yinoue@astro.isas.jaxa.jp}

\maketitle

%%%%%%%%%%
%%    Abstract    %%
%%%%%%%%%%
\begin{abstract}
The Fermi bubbles are gigantic gamma-ray structures in our Galaxy. The physical origin of the bubbles is still under debate. The leading scenarios can be divided into two categories. One is the nuclear star forming activity similar to extragalactic starburst galaxies and the other is the past active galactic nucleus (AGN) like activity of the Galactic center supermassive black hole. In this paper, we propose that metal abundance measurements will provide an important clue to probe their origin. Based on a simple spherically symmetric bubble model, we find that the generated metallicity and abundance pattern of the bubbles' gas strongly depend on assumed star formation or AGN activities. Star formation scenarios predict higher metallicities and abundance ratios of [O/Fe] and [Ne/Fe] than AGN scenarios do because of supernovae ejecta. Furthermore, the resultant abundance depends on the gamma-ray emission process because different mass injection histories are required for the different gamma-ray emission processes due to the acceleration and cooling time scales of non-thermal particles. Future X-ray missions such as {\it ASTRO-H} and {\it Athena} will give a clue to probe the origin of the bubbles through abundance measurements with their high energy resolution instruments.
\end{abstract}

%%%%%%%%%%%%
%%    Introduction    %%
%%%%%%%%%%%%
\section{Introduction}
\label{intro}
The Fermi bubbles are gigantic gamma-ray structures extending $\sim50^\circ$ north and south of the Galactic center (GC) with a longitudinal with $\sim40^\circ$ \citep{dob10,su10,ack14}. Structures roughly coincident with the gamma-ray bubbles are known in X-rays \citep{sno97,bla03}, microwave \citep{fin04,dob08,ade13}, and polarized radio \citep{car13}. Past activities of our Galaxy is believed to generate these structures. At lower latitudes $|b|\lesssim20^\circ$, an additional gamma-ray emission component is reported in the bubbles \citep{hop13}. This component would originate in millisecond pulsars or annihilation of dark matter particles rather than past activities of our Galaxy \citep{hop13}, although the latest analysis of the bubbles does not find that component because of a large systematic uncertainty \citep{ack14}.  

Gamma-ray emission of the bubbles is thought to be supplied by leptonic or hadronic processes, namely the inverse-Compton scattering of interstellar radiation field and the cosmic microwave background by electrons \citep[e.g.][]{che11,mer11,lac14}  or the hadronuclear process of protons (and ions) colliding with ambient gas in the bubbles \citep[e.g.][]{cro11,tho13,fuj13}. Both models can explain the microwave and gamma-ray data, although additional primary electrons or reacceleration of secondary leptons may be required in the hadronic scenario \citep{fuj14,ack14}.

In either case, the huge energy content of the bubbles an order of $10^{54-55}$~ergs \citep{su10,ack14} should be explained as well. A fundamental question on the bubbles is what powers the bubbles. Theoretically, two scenarios are proposed as the origin of the bubbles. Those are nuclear star-formation activity \citep[e.g.][]{cro11,car13,lac14} and past active galactic nucleus (AGN) activities of Sgr~A* \citep[e.g.][]{che11,zub11,guo12,mou14,yan13}. Although a jet-like structure in the bubbles was previously reported \citep{su12}, which supported the Sgr~A* jet scenario, that structure was not confirmed in the latest analysis \citep{ack14}. \citet{car13} has argued that the nuclear star formation activity scenario is favored based on the polarization measurement. However, the measured polarization features have been argued to be also reproduced by the AGN jet scenario \citep{yan13}. Other probes are necessary to investigate the origin of the bubbles.

\citet{kat13}, \citet{tah15}, and \citet{kat15} have recently carried out X-ray observations of the bubbles using the X-ray Imaging Spectrometers (XIS) onboard the {\it Suzaku} X-ray satellite. The observed diffuse X-ray emission shows the existence of $kT\simeq0.3$~keV thermal plasma which is slightly hotter than the surrounding Galactic Halo (GH) gas.  \cite{tah15} have further found the possible existence of 0.7~keV plasma is indicated in the northern cap region which is seen in the all sky map of the {\it Monitor of All-sky X-ray Image} (MAXI). They found the expansion velocity of the bubbles as $\sim300\ {\rm km\ s^{-1}}$ lower than most of previously proposed models \citep[e.g.][]{che11,zub11,mer11,guo12, lac14}. This velocity is supported by the measurement of the X-ray absorption line toward 3C~273 whose sightline passes through the neighborhood of the bubbles \citep{fan14}. Moreover, \citet{fox14} reported two high-velocity metal absorption components at -235 and +235~km/s using the spectrum of a quasar whose sightline passes through the bubbles.

In this paper, we propose X-ray abundance measurements in the bubbles will provide a unique key to identify their origin because the distributed elemental abundances depend on yields of ejecta and mass loading factor of ambient gas. The region of the bubbles was initially filled with the low metal GH gas. In the star forming activity scenarios, the bubbles are polluted by the elements produced by supernovae (SNe) whose abundances are different from that in the interstellar medium (ISM). On the other hand, in the AGN wind scenario, the abundance of the wind would be the same as the ambient ISM which accretes onto the Sgr~A*. The resultant abundance distribution in the bubbles is expected to be different between the AGN wind and star forming scenarios. We also argue prospect for future X-ray observations. We adopt solar abundances reported in \citet{asp09}. Thus, the solar metallicity is set to be $Z_\odot \simeq 0.0134$ rather than classical value of $Z_\odot\simeq 0.02$ \citep{and89}. 

In this Letter, we do not consider the the AGN jet scenario. The interior of the bubbles formed by jets would be polluted by metals of the jet itself. The Kelvin-Helmholtz instability is expected to be suppressed even with low level of viscosity \citep{guo12_2}. As the jets push the GH gas away, the metal mixing with the GH gas would not efficiently occur in the jet-induced bubbles. However, the jet composition is highly uncertain. Although pure pair jet models are excluded for  blazars \citep{sik00} and pairs may not survive the annihilation in the inner, compact and dense regions \citep{cel08}, there is still room for pairs in the jet, based on the energetics arguments \citep{sik05}. Moreover, iron emission lines are observed in the jet of the Galactic microquasar SS~433 \citep{mig02}. 

\section{Metal Enrichment in the Fermi Bubbles}
%%% WITHOUT JET %%%
\begin{table*}
\caption{Model Parameters for Metal-Enriched Outflows\label{tab:par}}
\begin{center}
\begin{tabular}{lllll}
\hline
\multicolumn{1}{l}{Origin} &  \multicolumn{2}{c}{Star formation}  & \multicolumn{1}{c}{AGN wind}& \multicolumn{1}{c}{AGN wind}\\
\multicolumn{1}{l}{Emission} & \multicolumn{1}{c}{Leptonic} & \multicolumn{1}{c}{Hadronic} & \multicolumn{1}{c}{Leptonic} & \multicolumn{1}{c}{Hadronic}  \\
\multicolumn{1}{l}{Reference} & \multicolumn{1}{c}{\citet{lac14}} & \multicolumn{1}{c}{\citet{cro14}} & \multicolumn{1}{c}{\citet{mou14}} & \multicolumn{1}{c}{\citet{zub11}}\\
\hline
SFR \ [$M_\odot/{\rm yr}$]  & 0.1 & 0.1 & - & - \\
IMF model & \citet{sal55} & \cite{kro01} & -& -\\
IMF ranges & 0.1-100 M$_\odot$ & 0.08-150 M$_\odot$ & - & - \\
 $\dot M_{\rm out}$\ [$M_\odot/{\rm yr}$] & 0.02 & 0.1 & 0.02 & 0.08$^{a}$ \\
$\beta$ & 2.0 & 6.3 & -$^{\rm b}$ & -$^{\rm b}$\\
\hline
 $Z_{\rm FB}/Z_{\odot}$ & 5.3$^{\rm c}$ & 2.2$^{\rm c}$  & 1.0$^{\rm c}$  & 0.45$^{\rm d}$\\
$X_{\rm Fe,FB}/X_{\rm Fe,\odot}$ & 2.3$^{\rm c}$  & 1.3$^{\rm c}$  & 1.0$^{\rm c}$  & 0.45$^{\rm d}$\\
{[O/Fe]} & 0.49$^{\rm c}$  & 0.30$^{\rm c}$  & 0.0$^{\rm c}$  & 0.0$^{\rm d}$\\
{[Ne/Fe]} & 0.58$^{\rm c}$  & 0.38$^{\rm c}$  & 0.0$^{\rm c}$  & 0.0$^{\rm d}$ \\
\hline
\multicolumn{5}{l}{$^{\rm a}$: This is required only for $\sim5\times10^4$~yr at $\sim6$~Myr ago \citep{zub11}.} \\
\multicolumn{5}{l}{$^{\rm b}$: $\beta$ does not affect results assuming $X_{i,{\rm ejecta}}=X_{i,{\rm ISM}}$ (see the details in the text).} \\
\multicolumn{5}{l}{$^{\rm c}$: Expected values behind the contact discontinuity, $R_{\rm cd}$. At larger radii, it will be the value of the GH gas.}\\
\multicolumn{5}{l}{$^{\rm d}$: Expected values in the bubbles elsewhere.}\\
 \end{tabular}
\end{center}
\end{table*}

To consider the metal enrichment in the bubbles, first we consider the metallicity in the outflow. We follow the descriptions in \citet{str09}, which discussed the outflow in the nearby starburst galaxy M~82. The net mass outflow rate from the GC is described as $\dot M_{\rm out} = \dot M_{\rm ejecta} + \dot M_{\rm ISM} \equiv \beta \dot M_{\rm ejecta}$, where $\dot M_{\rm ejecta}$ is the ejected mass outflow rate from the origin to the bubbles, $\dot M_{\rm ISM}$ is the loaded ISM mass rate, and $\beta$ is the mass loading factor. If $\beta=1$, no ISM gas is loaded. Given the star formation rate (SFR) and initial mass function (IMF), the ejected mass outflow rate is estimated. Then, by comparing with the required total mass outflow rate for the formation of the bubbles, the mass-loading factor is determined. For the nearby starburst galaxy M~82, the mass loading factor is in the range of $1.5\le\beta\le2.5$ \citep{str09}.

The elemental abundance $X_{i,\rm out}$ of an element $i$ in the outflow, i.e. the elemental mass fraction against the total baryons in the outflow gas, is
\begin{eqnarray}
\nonumber
X_{i, \rm out} &=& \frac{X_{i, \rm ejecta} \dot M_{\rm ejecta} + X_{i, \rm ISM} \dot M_{\rm ISM}}{\dot M_{\rm ejecta} + \dot M_{\rm ISM}}\\
&=&\frac{X_{i, \rm ejecta} + (\beta-1)X_{i, \rm ISM}}{\beta},
\label{eq:xi_out}
\end{eqnarray}
where $X_{i, \rm ejecta}$ and $X_{i, \rm ISM}$ is the abundance of the element in the ejecta and in the ambient ISM, respectively. Hereinafter, we assume $X_{i, \rm ISM}=X_{i,\odot}$ \citep[see e.g.][and references therein]{uchi13,nak13}.

Now we are interested in the abundance distribution in the bubbles. For the sake of simplicity, we assume a spherically symmetric bubble model. And, we also simply assume the GH gas had a distribution of $\rho_{\rm GH}\propto r^{-2}$ before the bubbles formed. Although the GH gas distribution has been under debate \citep[see e.g.][for details]{yao09,mil13,sak14}, recent measurements by {\it XMM-Newton} suggest $\rho_{\rm GH}\propto r^{-2.1}$ at $r\gtrsim0.35$~kpc \citep{mil13}. The adiabatic index of the gas is set to be $5/3$. We adopt the self-similar solution for the hydrodynamical evolution of the gas \citep[see e.g.][]{mih84,ost88}. Depending on the material injection history, the resulting matter distribution differs \citep[see e.g. Fig. 2 in][]{fuj13}. Instantaneous injection leads the gas mixing between outflow and the GH gas inside of the shock radius ($R_{\rm sh}$), while the continuous injection leads a compressed GH gas between the shock and the contact discontinuity at $R_{\rm cd}=0.84R_{\rm sh}$ and only outflow gas exists behind  $R_{\rm cd}$. The metal abundance in the bubbles will be given as follows. Instantaneous injection case gives
\begin{eqnarray}  
X_{i, \rm FB}
 &=\left\{\begin{array}{ll}
     \frac{X_{i, \rm out}\dot M_{\rm out}t_{\rm wind}+X_{i, \rm GH}M_{\rm GH}}{\dot M_{\rm out}t_{\rm wind}+ M_{\rm GH}}  &  (r \le R_{\rm sh})\\
    X_{i, \rm GH}& (r > R_{\rm sh}), 
    \end{array}\right.
    \label{eq:xi_fb_inst}
\end{eqnarray} 
while continuous injection case gives
\begin{eqnarray}  
X_{i, \rm FB}
 &=\left\{\begin{array}{ll}
     X_{i, \rm out} &  (r \le R_{\rm cd})\\
    X_{i, \rm GH}& (r > R_{\rm cd}), 
    \end{array}\right.
    \label{eq:xi_fb_cont}
\end{eqnarray} 
where $X_{i,\rm FB}$ is the abundance of an element $i$ in the bubbles, $t_{\rm wind}$ is the time scale where the wind is active, $X_{i,\rm GH}$ is the abundance of the element in the GH, and $M_{\rm GH}$ is the swept-up GH gas mass. The latter case is analogous to that in the wind of the starburst galaxies \citep[e.g.][]{str09}. We set $X_{i,\rm GH}=0.45X_{i,\odot}$ \citep{mil14}\footnote{We renormalize the reported value based on \citet{and89} to the latest solar abundance based on \citet{asp09}.}, although the GH gas metallicity is still uncertain \citep[see also][claiming solar metallicity]{sak14}.

From the {\it Suzaku} observations, the shock radius is indicated at around $10$~kpc from the GC \citep{kat13}. The swept-up halo gas mass is estimated as $\sim1.2\times10^8M_\odot$ using the spherical $\beta$ model \citep{mil13}, while we assume the gas distribution follows $r^{-2}$ for the simplicity. Once the abundances of the ejecta, the mass outflow rate, the timescale of wind activity and the mass loading factor are given, we can calculate the abundance distribution of the bubbles from Eqs. \ref{eq:xi_out}, \ref{eq:xi_fb_inst}, and \ref{eq:xi_fb_cont}. 

In the nuclear star formation scenario, stars distribute elements through SNe and stellar winds (SWs). We assume all stars have the solar abundances, since we assume $X_{i, \rm ISM}= X_{i,\odot}$. In this paper, we neglect the yields of SWs, which may be crucial for light elements. We do not discuss the H-burning products below. The contribution of SWs to yields of heavier elements is expected to be small for stars having solar abundances even taking into account rotation \citep[e.g.][]{hir05}. \citet{nom06} provide the yields from various mass core-collapse SNe and hypernovae (HNe) whose explosion energy is $\gtrsim10^{52}\ {\rm ergs}$ \citep{nom06}. The stars having mass of $\sim$25--140~$M_\odot$ in the main-sequence stage collapse to form a black hole. If the black hole has little angular momentum, little mass ejected. However, if the black hole rotates, the black hole eject matter through jet and it would be observed as a HN \citep{nom13}. 

We estimate the the SN ejecta abundances as follows \citep{nom06}. Given the IMF $\phi(M)dM$, the IMF-integrated yields normalized by the total mass of ejected materials are as follows \citep{nom06,tom07}\footnote{In \citet{nom06}, the IMF-integrated yields are normalized by the total amount of gases forming stars. Since we are interested in the abundance in the ejecta now, we adopt the Eq. \ref{eq:SN} in this Letter.}:
\begin{equation}
\label{eq:SN}
X_{i,{\rm ejecta}} = \frac{\int_{M_{\rm min}}^{M_{\rm max}}X_{i, {\rm SN}}(M_{\rm ej,SN}[M])M_{\rm ej,SN}(M)\phi(M)dM}{\int_{M_{\rm min}}^{M_{\rm max}}(M_{\rm ej,SN}[M]+ M_{\rm ej,SW}[M])\phi(M)dM},
\end{equation}
where $X_{i,{\rm ejecta}} $ is an integrated mass fraction of an element $i$, $X_{i, {\rm SN}}$ is mass fraction of $i$ linearly interpolated between nearest models of \citet{nom06} as a function of an ejected mass, $M_{\rm ej, SN}$ is an ejected mass by a SN, $M_{\rm ej, SW}$ is an ejected mass by SWs, and $M$ is the mass of a main sequence star. $M_{\rm min}$ and $M_{\rm max}$ is the minimum and maximum mass of stars, respectively. Following \citet{nom06}, We assume $M\le10M_\odot$ and $M\ge50 M_\odot$ stars do not yield any materials, i.e. $M_{\rm ej,SN}(M\le10M_\odot)=M_{\rm ej,SN}(\ge50 M_\odot)=0$. We assumed the fraction of HNe to whole SNe $\epsilon_{\rm HN}=0$ for $M<20M_\odot$ and $\epsilon_{\rm HN}=0.5$ for $M\ge20M_\odot$ \citep{kob06,nom06}.  

Following Eq. \ref{eq:SN}, $X_{\rm Fe, ejecta}$ for the Salpeter IMF with the mass range of 0.1--100~$M_\odot$ is 4.0~$X_{{\rm Fe, \odot}}$. In the nearby starburst galaxy M~82, its outflow is predicted to have $X_{\rm Fe, ejecta}\sim5X_{\rm Fe, \odot}$ \citep[see e.g.][]{str09}, although the assumed IMF and yields are different.

In the case of past AGN-like activities of Sgr~A*, the situation is different. The ejecta abundances reflect the accretion disk abundances which are the same as the ISM abundances. Thus, we set $X_{i,\rm ejecta}=X_{i,\rm ISM}=X_{i,\odot}$ in the AGN disk wind scenarios. Eq. \ref{eq:xi_out} implies that the yield of the outflow is $X_{i, \rm out}=X_{i,\odot}$. The mass loading factor does not affect results in the AGN disk wind scenarios.

In this paper, we consider the leptonic star formation (SF) scenario \citep[e.g.][]{lac14}, the hadronic SF scenario \citep[e.g.][]{cro11}, the leptonic AGN wind (AW) scenario \citep[e.g.][]{mou14}, and the hadronic AW scenario \citep[e.g.][]{zub11}. The model parameters are summarized in Table. \ref{tab:par}. As described below, we adopt the continuous injection case for the first three scenarios, while we adopt the instantaneous injection case for the hadronic AW scenario.

For the leptonic SF scenario, we adopt the fiducial model parameters in \citet{lac14}. They take the Salpeter initial mass function \citep{sal55} ranging 0.1--100~$M_\odot$ with the continuous SFR of 0.1~$M_\odot\ {\rm yr}^{-1}$. The mass outflow rate is 0.02~$M_\odot\ {\rm yr}^{-1}$ with $\beta$ of 2.0. 
\if0
However, the magnetic pressure becomes much greater than the external pressure with this fiducial SFR \citep[see \S.3.2. of][]{lac14} which leads different termination shock radius. To ease this tension, SFR is required to be $\sim0.2M_\odot\ {\rm yr}^{-1}$ \citep{lac14}, which produce more metals. However, this SFR is higher than the values reported in other studies on the SFR in the GC region \citep[see e.g.][]{mat11}. 
\fi

For the hadronic SF scenario, we adopt \citet{cro14} where they adopt the Kroupa initial mass function \citep{kro01} ranging 0.08--150~$M_\odot$ with the continuous SFR of 0.1~$M_\odot\ {\rm yr}^{-1}$ \citep{cro12}. The mass outflow rate is set to be 0.1~$M_\odot\ {\rm yr}^{-1}$. The mass-loading factor is estimated as follows. Given the SFR and IMF, the SN+SW ejected mass outflow rate is $0.016M_\odot\ {\rm yr}^{-1}$. Then, $\beta=\dot M_{\rm wind}/\dot M_{\rm ejecta}\simeq6.3$ assuming all the ejecta materials are injected into the bubbles \citep{cro12}.

For the leptonic AW model, we adopt the run~A of \citet{mou14}. They assume a radiative inefficiency accretion flow, but $2\times10^3$ times higher accretion rate than present value motivated by \citet{tot06} whose model can nicely explain various aspects of the GC observables by past Sgr~A* activity \citep[see][for details]{tot06}. The accretion disk wind has the continuous mass outflow for 12.3~Myr. 

For the hadronic AW model, we adopt \citet{zub11} which assume an Eddington accretion wind but blowing only for $t_{\rm wind}\sim5\times10^4$~yr at $\sim6$~Myr ago. The mass outflow rate from the GC region is terminated in other epochs. Since the mass injection occurs for short time scale comparing to the age of the bubble, the hadronic AW model can be regarded as the instantaneous injection. As described in \citet{zub11}, the mass outflow rate is $\sim8\times10^{-2}M_\odot {\rm yr}^{-1}$ during the Eddington phase.

\section{Results}
\label{sec:res}
The expected metallicity, iron abundance, and abundance ratios at a given radius are summarized in Table. \ref{tab:par}. We note that the observed values are integrated values on the line of sight as a function of the Galactic longitude and latitude. The metallicity in the bubbles will be $5.3~Z_\odot$, 2.2~$Z_\odot$, and $Z_\odot$ at $r \le R_{\rm cd}$ for the leptonic SF scenario, the hadronic SF scenario, and the leptonic AW scenario, respectively. At $r>R_{\rm cd}$, it will be the GH gas metallicity.  Therefore, as given in Eq. \ref{eq:xi_fb_cont}, the metallicity in the bubbles would have a clear jump at the contact discontinuity at $r\sim8$~kpc from the GC for the continuous injection cases. Because of the difference of the mass loading factor, the hadronic SF scenario predict lower metallicity than the leptonic SF scenario does. Since we assumed that the AGN disk wind and the loaded ISM have the solar abundance, the expected metallicity becomes $Z_\odot$. For the hadronic AW scenario, it will be kept at the GH gas metallicity level, 0.45~$Z_\odot$, at elsewhere. Although there is a small metallicity jump at the shock radius, that will be a factor of $\lesssim0.5$~\% jump. This is because the injected gas amount $\sim4.0\times10^3M_\odot$ is relatively smaller than the swept-up GH gas mass $\sim1.2\times10^8M_\odot$.

It is hard to distinguish models with current X-ray data through metallicities, since {\it Suzaku} data have huge uncertainties in deriving metallicities due to low photon statistics and its energy resolution. Further X-ray observations are required to unveil the origin of the bubbles through the abundance measurements. Interestingly, future missions such as {\it ASTRO-H} \citep{tak12} and {\it Athena} \citep{nan13} will have high energy-resolution spectrometers, which may enable us to study abundance ratios. Once elemental line emissions are clearly measured, we can reliably determine the metallicities and abundances in the bubbles. To compare with future data, we also evaluate the iron abundance and the abundance ratios which are the logarithm of the ratio of abundances compared to the solar abundance ratio. The iron abundance in the bubbles behind the contact discontinuity will be $2.3~X_{\rm Fe,\odot}$, $1.3~X_{\rm Fe,\odot}$, $X_{\rm Fe,\odot}$ for the leptonic SF scenario, the hadronic SF scenario, and the leptonic AW scenario, respectively. The iron abundance in the hadronic AW scenario will be 0.45~$X_{\rm Fe,\odot}$ at elsewhere. The abundance ratio of [O/Fe] behind the contact discontinuity will be 0.49, 0.30, and 0 for the leptonic SF scenario, the hadronic SF scenario, and the leptonic AW scenario, while it will be 0 for all models at larger radii.  It will also be zero at elsewhere for the hadronic AW scenario. We note that the solar abundance ratio corresponds to zero. Thus, AW scenarios give the value of zero. [Ne/Fe] also give the similar results as in [O/Fe], but [Ne/Fe] will be 0.58 and 0.38 for the leptonic SF scenario and the hadronic SF scenario behind the contact discontinuity.

\begin{figure}[t]
\centering
\includegraphics[width=7.2cm]{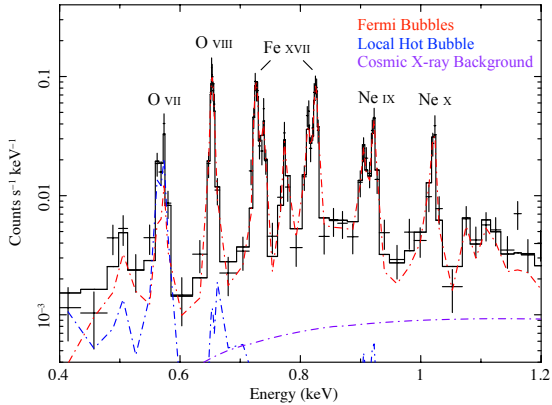} 
\caption{Simulated {\it ASTRO-H}/SXS spectrum of the Fermi bubbles with 200~ks exposure. The data represents the expected performance by SXS, while the black, red, blue, and purple curve represents contributions from all components, the Fermi bubbles, the local hot bubble, and the cosmic X-ray background. We assume the spectral parameters of the N-cap off region observed by {\it Suzaku} with the emission measure of $0.12\ {\rm cm^{-6}\ pc}$ \citep{tah15}, but we set the temperature of 0.3~keV and the metallicity of $0.45~Z_{\odot}$ for the bubbles.  [O/Fe], and [Ne/Fe] of the bubbles are set to be zero, i.e. the solar abundance ratios. If more metals exist in the bubbles, the stronger line emissions are expected. The position of the each line elements are indicated in the figure. \label{fig:sxs_metal} }
\end{figure}

We also perform spectral simulations for {\it ASTRO-H}\footnote{Response files are taken from \url{http://astro-h.isas.jaxa.jp/researchers/sim/response.html}. We adopt sxt-s\_120210\_ts02um\_of\_intallpxl.arf.gz for ARF, ah\_sxs\_7ev\_basefilt\_20090216.rmf.gz for RMF,  and sxs\_nxb\_7ev\_20110211\_1Gs.pha.gz for background  files.}. Figure. \ref{fig:sxs_metal} shows the simulated spectrum of the bubbles with 200~ks exposure for the Soft X-ray spectrometer (SXS) onboard {\it ASTRO-H}. Three components are included. Those are the bubbles, the local hot bubble, and the cosmic X-ray background following \citet{kat13,tah15}. Since the foreground GH gas component is not observed in the bubbles' region \citep{kat13,tah15,kat15}, the GH gas component is not included here. We assume the same spectral parameters of the N-cap off region observed by {\it Suzaku} with the emission measure of $0.12\ {\rm cm^{-6}\ pc}$ \citep{tah15} which is at the Galactic longitude of 355.5~deg and the Galactic latitude of 35.8~deg, but we set the temperature of 0.3~keV and the metallicity of $0.45~Z_{\odot}$ for the bubbles.  [O/Fe], and [Ne/Fe] of the bubbles are set to be zero, i.e. the solar abundance ratios. This situation roughly corresponds to the hadronic AW scenario. Under these assumptions, {\it ASTRO-H}/SXS can measure the metallicity of the bubbles as $Z_{\rm FB}=0.45^{+1.1}_{-0.21}Z_\odot$ and the abundance ratios as [O/Fe]=$0.00^{+0.16}_{-0.13}$ and [Ne/Fe]=$0.00^{+0.08}_{-0.11}$, where the errors represent 90\% confidence level. If more metals are contained, metallicity and abundance ratios are more precisely constrained because of stronger line fluxes. Although precise determination of the metallicity is hard, we can determine abundance ratios precisely through the {\it ASTRO-H} observations. If {\it ASTRO-H}/SXS observe higher abundance ratios, it would strongly support the star forming activity scenarios as the origin of the bubbles. Moreover, precise determination of the abundance ratios will help us to distinguish the gamma-ray emission process of the bubbles.

\section{Discussion and Conclusion}
In this paper, we showed that measurements of abundances in the bubbles will provide a unique clue to unveil their origin. The metal enrichment in the bubbles strongly depends on the bubbles formation scenarios and their emission mechanisms. It is still hard to determine the metallicities or abundances of the bubbles with current X-ray instruments. Further data or future missions are required. {\it ASTRO-H}/SXS can achieve a factor of 10--100 times better energy resolution than {\it Suzaku}/XIS do. Such high energy resolution will allow us to determine lines and their ratios. Based on the spectral simulation analysis, {\it ASTRO-H}/SXS will clearly detect line emissions. If high abundance ratios are obtained by {\it ASTRO-H}/SXS measurements, it will strongly support the star forming scenario as the origin of the bubbles. Moreover, precise measurement of the abundance ratios will enable us to investigate the gamma-ray emission process. Furthermore, future X-ray mission {\it Athena} \citep{nan13} will have similar instrument but with higher energy resolution and larger effective area. These future X-ray missions will enable us to understand the origin of the bubbles through the elemental abundances in the bubbles. 

% Thermal Conduction
For continuous injection models, we do not take into account the thermal conduction effect. As hot outflow gas exists behind the contact discontinuity, compressed gas can be heated up by the thermal conduction and flow behind the contact discontinuity. The abundance of the gas behind the contact discontinuity would be smaller than estimated. The thermal conduction time scale is given as $t_{\rm cond}\simeq10^8 ({n}/{4\times10^{-3}\ {\rm cm^{-3}}})({l_T}/{1.6\ {\rm kpc}})^2 ({kT}/{0.3\ {\rm keV}})^{-5/2} \ {\rm yr}$ \citep{kaw02}, where $n$ is the gas density taken from \citet{kat13}, $l_T$ is the thermal conduction length assumed to be the thickness of the compressed region, and $kT$ is gas temperature set to be 0.3~keV \citep{kat13}. Since the age of the bubble is expected to be in the order of 10~Myr for the leptonic SF and leptonic AW scenarios, the results will not significantly change. However, in the case of the hadronic SF scenarios, the age would be comparable to the thermal conduction time scale. The actual abundance would be lower than that estimated in this paper.

We assumed that the interior of the bubbles is described by a single temperature. In nearby starburst galaxies, observed X-ray emitting gas is composed of multi-temperature plasma \citep{str02}. Single temperature modelling may result in erroneous abundance measurement. Here, the physical scale of the observed regions of the nearby starburst galaxies extends to $\sim3$~kpc \citep{str02}, while that scale of the Field-of-View (FoV) of {\it Suzaku}/XIS and {\it ASTRO-H}/SXS at the GC is $\sim40$~pc and $\sim7$~pc, respectively. The expected $t_{\rm cond}$ in the observable regions of the bubbles by {\it Suzaku}/XIS and {\it ASTRO-H}/SXS becomes much shorter than the age of the bubbles. Thus, single temperature models work for the bubbles for pointing X-ray observations. Furthermore, the current X-ray spectra of the bubbles are well described by a single temperature model \citep{kat13,tah15,kat15}, although stacking analysis of the northern cap region indicates possible existence of another 0.7~keV plasma \citep{tah15}. With {\it ASTRO-H}/SXS, we can observationally distinguish another temperature component by comparing the temperature based on single temperature spectral fit and that based on line ratios in each field.

Non-thermal X-ray emission may underlie the thermal component as non-thermal emission is observed in radio and gamma-ray. Significant contribution of non-thermal emission may be crucial for deriving abundances. \citet{kat13} observationally constrained the non-thermal flux associated with the bubbles as $<9.3\times10^{-9}\ {\rm erg\ cm^{-2}\ s^{-1}\ sr^{-1}}$ in the 2--10~keV energy range, which is negligible comparing to the observed thermal flux. Theoretically, non-thermal X-ray flux of the bubbles is expected to be less than the observational upper limit through multi-wavelength spectral modelling \citep[see e.g.][]{kat13,ack14,fuj14}. 

% Geometry
\if0
For the geometry, we assumed a simple spherically symmetric model. The observed gamma-ray structures show two bubbles above and below the Galactic disk. Since we simply assume a spherical model, the actual abundance would be higher because the swept-up gas in the two bubbles case mass must be smaller than the spherical case. To reproduce the observed shape, we need to take into account the gas distribution in the GC region where the dense gas exist in the Galactic plane. 
\fi
% SFR and SNe Ia
We do not take into account the yields of Type Ia supernovae (SNe~Ia) considering the uncertainties of the SNe~Ia rate in the GC which is not observationally well constrained. SNe~Ia are the thermonuclear explosions of accreting white dwarfs and produce Fe and little $\alpha$--elements \citep[e.g.][]{iwa99}. It is known that the cosmic SNe~Ia rate is a factor of 3--10 lower than the cosmic core-collapse SNe rate \citep{hor10,hor11}. SN~Ia explosion occurs not simultaneously with star formation but delays. Delay time distribution (DTD) of SNe~Ia is represented by a power-law form \citep[see e.g.][]{tot08}. By assuming a constant star formation history (SFH) and a power-law DTD, the expected SNe~Ia rate is $\sim0.01$ per century which is roughly consistent with the estimate of $0.03\pm0.02$ per century \citep{sch07} based on the empirical relation between the rate and the stellar mass \citep{man05}. The resultant iron abundance increases by 2\% and 20\% for the leptonic SF scenario \citep{lac14} and  the hadronic SF scenario \citep{cro12}, respectively. We adopted the W7 model in \citet{iwa99} for the yields of SNe~Ia and a power-law DTD following \citet{yat13}. However, the SFRs in the nuclear bulge at $\gtrsim30$--70~Myr ago were about an order of magnitude lower than that at $\sim1$~Myr ago \citep{mat11}. Taking into account this SFH, the iron abundance does not change for the leptonic SF scenario, while it increases 2\% for the hadronic SF scenario. Considering the uncertainties of future {\it ASTRO-H}/SXS measurements (see \S. \ref{sec:res}), the metal enrichment by SNe~Ia in the bubbles would be negligible comparing to abundance measurement uncertainties.

Abundances of stars and ISM in the GC region are assumed to be the solar. However, those abundances in the GC are still under debate. Various observations suggest that the GC metallicity is at least in the range of $Z_\odot\lesssim Z_{\rm GC} \lesssim 2 Z_\odot$  \citep[see the appendix A of][for details]{cro12}, although their elemental abundances are uncertain. If we assume $2Z_{\odot}$ for ISM and stars in the GC, the resulting metallicity behind $R_{\rm cd}$ in the star formation scenarios increases by $\sim$20\% comparing to the case with solar abundance progenitor stars. We adopt the yields described in \citet{por98} which give the yields for stars having up to $2.5Z_{\odot}$, while the yields for stars having $>Z_\odot$ are not given in \citet{nom06}. However, the effect of metal enrichment from HNs are not included in this comparison since those are not provided in \citet{por98}.

% Neutrino
\if0
Neutrinos from the bubbles will be also a key to probe the emission mechanism. If the gamma-ray emission of the bubbles originates in $pp$ collisions, we can speculate the expected neutrino flux \citep[e.g.][]{ahl14}. Based on the latest measurement of the bubbles \citep{ack14}, the expected neutrino flux is below the IceCube neutrino flux level \citep{ahl14}. Further neutrino data may be required to distinguish the emission process in the bubbles from the multimessenger point of view.
\fi
% Rotation Measure?

\bigskip
We thank the anonymous referee for useful comments and suggestions. We also thank to Noriko Yamasaki and Dmitry Malyshev for useful comments and discussions. Y.I. acknowledges support by the JAXA international top young fellowship.

\end{document}